# Sliding of a liquid spherical droplet in an external insoluble liquid at low Reynolds numbers

Peter Lebedev-Stepanov

Shubnikov Institute of Crystallography, Kurchatov Complex of Crystallography and Photonics, Leninskiy Prospekt 59, Moscow 119333, Russia

E-mail: lebstep.p@crys.ras.ru

The experiment shows that small liquid droplets under the action of gravity and the Archimedes force move in the external viscous liquid practically according to the Stokes drag force equation, and not in accordance with the Hadamard–Rybczynski (HR) formula, which was specially developed to describe the motion of a liquid droplet in an external viscous liquid. Various mechanisms are proposed to explain this: increased viscosity at the interface between two liquids and the presence of unaccounted surfactants. However, there is another fundamental mechanism that has not been taken into account. It can be expected that the velocities of such liquids, insoluble in each other, may not equalize at the boundary of the droplet. No slip condition may be may be unnatural at the droplet interface. In this paper, the Navier condition is applied to the liquid-liquid boundary for the first time. A generalized HR equation is obtained. If slip length $\lambda=0$ that equation transforms into the usual HR equation. At certain $\lambda$, we arrive at a model with continuity of the components of the viscous stress tensor at the interface of two fluids. For infinite viscosity of the drop, it becomes a well-known relation generalizing the Stokes drag force for a solid sphere, taking into account the boundary condition of partial slip

## 1. Introduction

The experiment shows that small liquid droplets under the action of gravity and the Archimedes force move in the external viscous liquid practically according to the Stokes law (i.e., Stokes drag force formula), and not in accordance with the Hadamard–Rybczynski equation, which was specially developed to describe the motion of a liquid droplet in an external viscous liquid [1,2]. Various mechanisms are proposed to explain this: increased viscosity at the interface between two liquids and the presence of unaccounted surfactants [2,3]. The latter method has gained dominant importance in interpreting the experiments.

We believe that in reality many factors can affect the result, so the role of the surfactant unaccounted for in the Hadamard and Rybczynski model cannot be denied, as shown by many researchers. However, there is another fundamental mechanism that has not been taken into account. Namely, that studies assume that there is no discontinuity in the velocities of the internal and external liquids at the interface (no slip condition). However, since it is assumed that the liquids do not mix, repel each other, and therefore have low affinity, it is very likely that the nature of their interaction at the interface can be described as sliding.

Thus, the formulation of boundary conditions in deriving the Hadamard–Rybczynski formula is debatable. There is no slip condition that links the velocities of the external and internal liquids on the surface of the drop. Since we are considering immiscible liquids with a clear interface, it can be expected that the velocities of such liquids, insoluble in each other, may not equalize at the boundary of the droplet.

One way to circumvent the condition of no sliding at the interface of two liquids is to replace it with the requirement of continuity of all components of the viscous stress tensor, including $\varphi\varphi$- and $\theta\theta$-components which would be natural if we are talking about a liquid without discontinuities and oscillations. It is shown here that such a model has a very narrow applicability in describing the experiment.

Another way is a model with arbitrary slip length at liquid-liquid interface. In this paper, the Navier condition is applied to the liquid-liquid boundary for the first time. A generalized Hadamard-Rybczynski equation is obtained.

It is shown that the model of partial slip at a liquid-liquid boundary is similar from a mathematical point of view to the Boussinesq model [1,10], although not entirely identical to it.

**2. Derivation of the Hadamard–Rybczynski formula from the general solution of the axisymmetric Stokes equations**

Let us consider a liquid droplet placed inside another liquid. Both liquids are insoluble in each other, do not mix with each other, and have a clear interface. The drop has a spherical shape stabilized by interfacial surface tension. The *z* axis is oriented vertically upwards (Fig. 1). If the drop density $\rho'$ is less than the liquid density $\rho$, it floats vertically upwards with a steady-state velocity $V_0$.

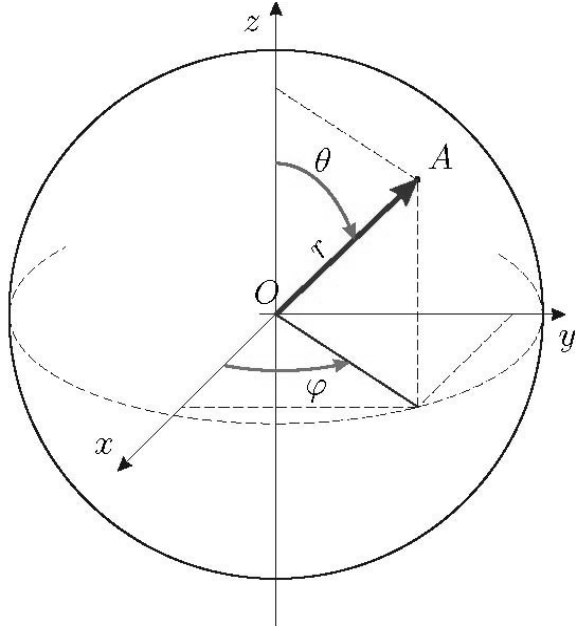

**Fig. 1.** Spherical coordinates $(r, \theta, \varphi)$, which include the radial coordinate, polar and azimuthal angles, respectively.

The hydrostatic pressure in the external liquid on the surface of the drop is determined by the formula

$$p_0 = -\rho g(h - R\cos\theta), \quad (1)$$

where $h$ is the depth of immersion of the center of the drop. The minus sign before the expression on the right takes into account that the hydrostatic pressure is compressive.

Similarly, the internal liquid creates pressure:

$$p_0' = \rho' g R\cos\theta + c, \quad (2)$$

where $c$ is a constant.

The surface of the liquid sphere is affected by the resulting pressure equal to the difference between Eq. (2) and Eq.(1)

$$\Delta p_0(\theta) = p_0 - p_0' = (\rho - \rho')gR\cos\theta, \quad (3)$$

where we have neglected the terms independent of the polar angle. It determines the Archimedean force acting on the drop:

$$F_z = 2\pi R^2 \int_0^\pi \Delta p_0(\theta) \sin\theta \cos\theta \, d\theta = (\rho - \rho')gv, \quad (4)$$

where $v = \dfrac{4}{3}\pi R^3$ is the volume of the liquid sphere.

The constancy of the steady-state velocity of the drop is ensured by the resistance force, which depends on the velocity of the drop, is equal in magnitude to the force (4) and is directed in the opposite direction. The resistance force is associated with the steady-state flow of the external liquid around the drop and the flow of another liquid inside the drop.

In the frame of reference associated with the droplet, the external liquid at $r \to \infty$ satisfies the boundary conditions

$$V_r(\infty, \theta) = V_0 \cos\theta, \tag{5}$$

$$V_\theta(\infty, \theta) = -V_0 \sin\theta. \tag{6}$$

The boundary conditions on the droplet surface are:

$$V_r(R, \theta) = 0, \tag{7}$$

$$V_r'(R, \theta) = 0, \tag{8}$$

which means that the droplet does not change its spherical shape (the shaded projection of the velocity, like all the shaded quantities below, characterizes the liquid inside the droplet).

As for the polar component of the velocity on the droplet, in this case the condition of no slip of one liquid over the other is usually accepted [1,4,5]:

$$V_\theta(R, \theta) = V_\theta'(R, \theta). \tag{9}$$

In addition, the continuity of viscous stresses acting normally on the interface between two fluids is used as boundary conditions: In the framework of an axisymmetric problem, these stresses have the form [1]:

$$\sigma_{rr} = \sigma_{rr}', \tag{10}$$

which in expanded form means

$$-(p + p_0) + 2\eta \frac{\partial V_r}{\partial r} = -(p' + p_0') + 2\eta' \frac{\partial V_r'}{\partial r} \tag{11}$$

and

$$\sigma_{r\theta} = \sigma_{r\theta}', \tag{12}$$

i.e.

$$\eta \left( \frac{1}{r} \frac{\partial V_r}{\partial \theta} + \frac{\partial V_\theta}{\partial r} - \frac{V_\theta}{r} \right) = \eta' \left( \frac{1}{r} \frac{\partial V_r'}{\partial \theta} + \frac{\partial V_\theta'}{\partial r} - \frac{V_\theta'}{r} \right). \tag{13}$$

Here $p$ and $p'$ denote the hydrodynamic pressures in the outer and inner fluids, respectively.

Equation (11) can be conveniently rewritten taking into account (1)-(3) as

$$-p + 2\eta \frac{\partial V_r}{\partial r} + p' - 2\eta' \frac{\partial V_r'}{\partial r} = p_0 - p_0' = (\rho - \rho')gR \cos\theta. \tag{14}$$

Similarly to the cases considered above, axisymmetric boundary conditions select those modes of the general solution that must be involved. All other modes can satisfy the boundary conditions only with zero coefficients. Thus, we will seek a solution to the axisymmetric problem.

The general solution of the axisymmetric problem for the Stokes equations in a spherical coordinate system is presented in Table 1. The derivation of the formulas is given in Ref. [6].

We see that the boundary conditions (5)-(6) and (4) require solutions containing $\cos\theta$ and $\sin\theta$. The solutions for $V_r$ and $p$ have the form of a series expansion in Legendre polynomials, and the solution for $V_\theta$ is represented by a series in the adjoint Legendre polynomial $P_l^1$. The orthogonality of the polynomials suggests that in order to satisfy the boundary conditions, it is necessary to limit ourselves to the terms of the series that contain $\cos\theta$ or $\sin\theta$. The remaining terms must be discarded, since the boundary conditions do not allow nonzero solutions corresponding to these excess terms to exist. Thus, in the solutions under consideration, we leave only the terms with $l=1$.

Table 1. Solutions of internal and external axisymmetric problems in a spherical coordinate system obtained in the representation of a vector potential. The radial and polar components of the incompressible fluid velocity, pressure and stream function are shown in the rows from top to bottom; $P_l(\cos\theta)$ is the Legendre polynomial, $P_l^1(\cos\theta)$ is the attached Legendre function of the first order.

| Internal problem | External problem |
|---|---|
| $V_r(r,\theta) = \sum_{l=1}^{\infty} l(l+1)\left\{\dfrac{a_l}{4l+6}\left(\dfrac{r}{R}\right)^{l+1} + c_l\left(\dfrac{r}{R}\right)^{l-1}\right\} P_l(\cos\theta)$ | $V_r(r,\theta) = -\sum_{l=1}^{\infty} l(l+1)\left\{\dfrac{b_l}{4l-2}\left(\dfrac{R}{r}\right)^{l} + d_l\left(\dfrac{R}{r}\right)^{l+2}\right\} P_l(\cos\theta) + d_0\left(\dfrac{R}{r}\right)^{2}$ |
| $V_\theta(r,\theta) = \sum_{l=1}^{\infty}\left\{a_l\dfrac{l+3}{4l+6}\left(\dfrac{r}{R}\right)^{l+1} + c_l(l+1)\left(\dfrac{r}{R}\right)^{l-1}\right\} P_l^1(\cos\theta)$ | $V_\theta(r,\theta) = \sum_{l=1}^{\infty}\left\{b_l\dfrac{l-2}{4l-2}\left(\dfrac{R}{r}\right)^{l} + d_l l\left(\dfrac{R}{r}\right)^{l+2}\right\} P_l^1(\cos\theta)$ |
| $p(r,\theta) = \dfrac{\eta}{R}\sum_{l=0}^{\infty}(l+1)a_l\left(\dfrac{r}{R}\right)^{l} P_l(\cos\theta)$ | $p(r,\theta) = -\dfrac{\eta}{R}\sum_{l=1}^{\infty} l b_l\left(\dfrac{R}{r}\right)^{l+1} P_l(\cos\theta)$ |
| $\Psi(r,\theta) = -R^2\sin\theta\sum_{l=1}^{\infty}\left\{\dfrac{a_l}{4l+6}\left(\dfrac{r}{R}\right)^{l+3} + c_l\left(\dfrac{r}{R}\right)^{l+1}\right\} P_l^1(\cos\theta)$ | $\Psi(r,\theta) = R^2\sin\theta\sum_{l=1}^{\infty}\left\{\dfrac{b_l}{4l-2}\left(\dfrac{R}{r}\right)^{l-2} + d_l\left(\dfrac{R}{r}\right)^{l}\right\} P_l^1(\cos\theta) - d_0 R^2\cos\theta$ |

For the internal problem, we have

$$V_r = 2\left\{\frac{a}{10}\left(\frac{r}{R}\right)^2 + c\right\} P_1, \tag{15}$$

$$V_\theta = \left\{a\frac{4}{10}\left(\frac{r}{R}\right)^2 + 2c\right\} P_1^1. \tag{16}$$

$$p = p_0 + \frac{2a\eta r}{R^2} P_1. \tag{17}$$

For the external problem, we obtain

$$V_r = -2\left\{\frac{b}{2}\frac{R}{r} + d\left(\frac{R}{r}\right)^3\right\}\cos\theta, \qquad (18)$$

$$V_\theta = \left\{\frac{bR}{2r} - d\left(\frac{R}{r}\right)^3\right\}\sin\theta, \qquad (19)$$

$$p = -\frac{\eta}{R}b\left(\frac{R}{r}\right)^2\cos\theta. \qquad (20)$$

The external fluid must satisfy the conditions at infinity. In this case, it is necessary to use the solution of the internal problem (15)-(16) for $a = 0$. Indeed, this will allow us to satisfy conditions (5)-(6) if all other terms of the solution vanish due to the presence of $r$ in the denominator. Substituting (15)-(16) for a=0 into (5)-(6) yields

$$2c = V_0. \qquad (21)$$

Adding the solution of the external problem here, we find that the external fluid is described by the equations

$$V_r = \left\{V_0 - b\frac{R}{r} - 2d\left(\frac{R}{r}\right)^3\right\}\cos\theta, \qquad (22)$$

$$V_\theta = \left\{\frac{bR}{2r} - d\left(\frac{R}{r}\right)^3 - V_0\right\}\sin\theta, \qquad (23)$$

$$p = -\frac{\eta}{R}b\left(\frac{R}{r}\right)^2\cos\theta. \qquad (24)$$

In order to have bounded solutions inside the droplet, the internal fluid must be described only by the solution of the internal problem

$$V_r' = 2\left\{\frac{a'}{10}\left(\frac{r}{R}\right)^2 + c'\right\}\cos\theta, \qquad (25)$$

$$V_\theta' = -\left\{a'\frac{4}{10}\left(\frac{r}{R}\right)^2 + 2c'\right\}\sin\theta. \qquad (26)$$

$$p' = p_0' + \frac{2a'\eta' r}{R^2}\cos\theta. \qquad (27)$$

Let us to substitute the obtained relations into the boundary conditions on the droplet surface. For the radial component of the velocity of the external fluid (22) we have

$$V_r(R,\theta) = (V_0 - b - 2d)\cos\theta = 0. \qquad (28)$$

Hence

$$b = V_0 - 2d. \qquad (29)$$

Then, substituting (19) into (18)-(20), we obtain

$$V_r = \left\{V_0\left(1-\frac{R}{r}\right)+2d\left[\frac{R}{r}-\left(\frac{R}{r}\right)^3\right]\right\}\cos\theta, \qquad (30)$$

$$V_\theta = \left\{V_0\left(\frac{R}{2r}-1\right)-d\left[\frac{R}{r}+\left(\frac{R}{r}\right)^3\right]\right\}\sin\theta, \qquad (31)$$

$$p = -\frac{\eta}{R}(V_0-2d)\left(\frac{R}{r}\right)^2\cos\theta. \qquad (32)$$

Similarly, for the radial component of the internal fluid velocity (15) we have

$$V_r'(R,\theta) = 2\left\{\frac{a'}{10}+c'\right\}\cos\theta = 0, \qquad (33)$$

hence

$$a' = -10c'. \qquad (34)$$

Therefore

$$V_r' = 2c'\left\{1-\left(\frac{r}{R}\right)^2\right\}\cos\theta, \qquad (35)$$

$$V_\theta' = 2c'\left\{2\left(\frac{r}{R}\right)^2-1\right\}\sin\theta. \qquad (36)$$

$$p' = -\frac{20c'\eta r}{R^2}\cos\theta. \qquad (37)$$

The boundary condition (9), meaning the absence of slippage (no slip), gives

$$-\frac{1}{2}V_0-2d = 2c'. \qquad (38)$$

Then equations (35)-(37) are rewritten as

$$V_r' = \left(\frac{1}{2}V_0+2d\right)\left\{\left(\frac{r}{R}\right)^2-1\right\}\cos\theta, \qquad (39)$$

$$V_\theta' = \left(\frac{1}{2}V_0+2d\right)\left\{1-2\left(\frac{r}{R}\right)^2\right\}\sin\theta. \qquad (40)$$

$$p' = (5V_0+20d)\frac{\eta' r}{R^2}\cos\theta. \qquad (41)$$

Substituting the components of velocity and hydrodynamic pressures into condition (11) taking into account that on the sphere

$$-p+2\eta\frac{\partial V_r}{\partial r} = \frac{3\eta}{R}(V_0+2d)\cos\theta, \qquad (42)$$

$$-p'+2\eta'\frac{\partial V_r'}{\partial r} = -(V_0+4d)\frac{3\eta'}{R}\cos\theta \tag{43}$$

and further, substituting into (14), we obtain

$$(\eta+\eta')V_0 = \frac{1}{3}(\rho-\rho')gR^2 - (2\eta+4\eta')d. \tag{44}$$

Similarly, condition (13) gives

$$2\eta d = -\eta'\left(\frac{1}{2}V_0 + 2d\right), \tag{45}$$

or

$$d = -\frac{V_0\eta'}{4(\eta+\eta')}. \tag{46}$$

Substituting Eq. (46) into Eq. (44), we obtain the well-known Hadamard–Rybczynski equation [1]

$$V_0 = \frac{2(\eta+\eta')(\rho-\rho')gR^2}{3\eta(2\eta+3\eta')}. \tag{47}$$

Note that the z-axis is directed upward, so that the velocity (47), as expected, is positive if the drop density is less than the density of the external liquid.

Taking into account that the drop is acted upon by an external force (4), this expression can be rewritten as

$$V_0 = \frac{(\eta+\eta')F}{2\pi\eta(2\eta+3\eta')R}. \tag{48}$$

In particular, if the viscosity of the liquid inside the drop is much greater than that of the external liquid, we have

$$V_0 \approx \frac{2(\rho-\rho')gR^2}{9\eta}. \tag{49}$$

or

$$V_0 \approx \frac{F}{6\pi\eta R}, \tag{50}$$

which is identical to the Stokes drug force for a solid spherical particle.

### 3. Model with continuous viscous stress tensor

The formulation of boundary conditions in deriving the Hadamard–Rybczynski formula is debatable. Let us consider no slip condition (9), which links the velocities of the external and internal liquids on the surface of the drop. Since we are considering immiscible liquids with a clear interface, it can be expected that the velocities of such liquids, insoluble in each other, may not equalize at the boundary of the droplet.

Similarly, when water flows around a hydrophobic surface, sliding without sticking is allowed. Let us see what condition can replace (9) in the case of sliding of a liquid sphere.

The difference between a solid sphere sliding in an external liquid and a liquid drop is that at the boundary with a solid body the liquid ends and viscous stresses are compensated by stresses of a different nature. In the case of a liquid-liquid boundary, it is natural to require the continuity of all components of the viscous stress tensor at the interface. This is considered in Ref. [5].

If the considered linearized stationary equation does not allow for the stitching of all components of the viscous stress tensor at the boundary of two liquids, then, generally speaking, we have a nonequilibrium picture that must be described by a non-stationary equation.

In the framework of the axisymmetric problem considered above, the continuities of only two non-zero components of the viscous stress tensor were ensured, $\sigma_{rr}$ and $\sigma_{r\theta}$, according to conditions (10) and (12). When deriving the Hadamard formula, only the listed stresses are used, which transmit the force through an area normal to the interface surface of the liquids.

But there are two more non-zero stresses that do not vanish in the axisymmetric problem, which act tangentially to the spherical surface of the droplet [4,5], namely:

$$\sigma_{\theta\theta} = -p - p_0 + 2\eta\left(\frac{1}{r}\frac{\partial V_\theta}{\partial \theta} + \frac{V_r}{r}\right), \quad (51)$$

$$\sigma_{\varphi\varphi} = -p - p_0 + 2\eta\left(\frac{V_r}{r} + \frac{V_\theta \cot\theta}{r}\right). \quad (52)$$

and similar tensions in the liquid inside the drop.

If these stress tensor components in the external and internal liquids are not stitched together, instability and turbulence will arise. This goes beyond the description using stationary equations. Indeed, let us imagine that the interface between two liquids experiences small oscillations, i.e. surface waves. In this case, local deviations of the normal to the area from the radial direction occur, i.e. non-zero projections in the direction of the polar and axial axes, so that forces corresponding to stresses given by Eqs. (51) and (52) are transmitted to the interface, which can amplify surface waves, creating a turbulent non-stationary picture, the description of which goes beyond the stationary model under consideration.

So, let us consider the stitching conditions:

$$\sigma_{\theta\theta} = \sigma_{\theta\theta}', \quad (53)$$

$$\sigma_{\varphi\varphi} = \sigma_{\varphi\varphi}'. \quad (54)$$

Conditions (53)-(54) can be rewritten in a form similar to Eq. (14):

$$-p + 2\eta\left(\frac{1}{R}\frac{\partial V_\theta}{\partial \theta} + \frac{V_r}{R}\right) + p' - 2\eta'\left(\frac{1}{R}\frac{\partial V_\theta'}{\partial \theta} + \frac{V_r'}{R}\right) = p_0 - p_0' = (\rho - \rho')gR\cos\theta, \quad (55)$$

$$-p + 2\eta\left(\frac{V_r}{R} + \frac{V_\theta \cot\theta}{R}\right) + p' - 2\eta'\left(\frac{V_r'}{R} + \frac{V_\theta' \cot\theta}{R}\right) = p_0 - p_0' = (\rho - \rho')gR\cos\theta. \tag{56}$$

Subtracting Eq. (56) from Eq. (55), we obtain:

$$2\eta\left(\frac{1}{R}\frac{\partial V_\theta}{\partial \theta} + \frac{V_r}{R}\right) - 2\eta\left(\frac{V_r}{R} + \frac{V_\theta \cot\theta}{R}\right) - 2\eta'\left(\frac{1}{R}\frac{\partial V_\theta'}{\partial \theta} + \frac{V_r'}{R}\right) + 2\eta'\left(\frac{V_r'}{R} + \frac{V_\theta' \cot\theta}{R}\right) = 0 \tag{57}$$

or, after transformations:

$$\eta\left(\frac{\partial V_\theta}{\partial \theta} - V_\theta \cot\theta\right) = \eta'\left(\frac{\partial V_\theta'}{\partial \theta} - V_\theta' \cot\theta\right). \tag{58}$$

Similarly to the derivation of the Hadamard–Rybczynski formula, we use the general solution presented in Table 1, taking into account that the boundary conditions require solutions containing $\cos\theta$ and $\sin\theta$, i.e. corresponding to $l=1$. For the internal problem we have solutions given by Eqs. (15)-(17), and for the external problem we have Eqs. (18)-(20).

Condition (58) within the framework of an axisymmetric problem can be transformed to the form [7]:

$$\frac{\partial P_l^1}{\partial \theta} - P_l^1 \cot\theta = -\sin\theta \frac{\partial P_l^1}{\partial \cos\theta} - \frac{\sin\theta}{\cos\theta}P_l^1 = -\frac{1}{2}l(l+1)P_l + \frac{1}{2}P_l^2 + \frac{1}{2}P_l^2 + \frac{1}{2}l(l+1)P_l = P_l^2. \tag{59}$$

In particular, for $l=1$ we have $P_1^2 \equiv 0$, so that conditions (55) and (56) identically coincide.

The external fluid must satisfy the conditions at infinity. In this case, it is necessary to use the solution of the internal problem given by Eqs. (15)-(17) for $a=0$. Indeed, this will allow us to satisfy conditions (5)-(6). Substituting Eqs. (15)-(17) for $a=0$ into Eqs. (5)-(6) gives

$$2c = V_0. \tag{60}$$

Adding here the solution of the external problem, we obtain that the external fluid is described by the equations

$$V_r = \left\{V_0 - b\frac{R}{r} - 2d\left(\frac{R}{r}\right)^3\right\}\cos\theta, \tag{61}$$

$$V_\theta = \left\{\frac{bR}{2r} - d\left(\frac{R}{r}\right)^3 - V_0\right\}\sin\theta, \tag{62}$$

$$p = -\frac{\eta}{R}b\left(\frac{R}{r}\right)^2\cos\theta. \tag{63}$$

In order to have bounded solutions inside the droplet, the internal fluid must be described only by the solution of the internal problem

$$V_r' = 2\left\{\frac{a'}{10}\left(\frac{r}{R}\right)^2 + c'\right\}\cos\theta, \tag{64}$$

$$V_\theta' = -\left\{a'\frac{4}{10}\left(\frac{r}{R}\right)^2 + 2c'\right\}\sin\theta. \tag{65}$$

$$p' = p_0' + \frac{2a'\eta' r}{R^2}\cos\theta. \tag{66}$$

We substitute the obtained relations into the boundary conditions on the droplet surface. For the radial component of the velocity of the external fluid (61) we have

$$V_r(R,\theta) = (V_0 - b - 2d)\cos\theta = 0. \tag{67}$$

Hence

$$b = V_0 - 2d. \tag{68}$$

Then, substituting Eq. (68) into Eqs. (61)-(63), we obtain the equations describing the external fluid

$$V_r = \left\{V_0\left(1 - \frac{R}{r}\right) + 2d\left[\frac{R}{r} - \left(\frac{R}{r}\right)^3\right]\right\}\cos\theta, \tag{69}$$

$$V_\theta = \left\{V_0\left(\frac{R}{2r} - 1\right) - d\left[\frac{R}{r} + \left(\frac{R}{r}\right)^3\right]\right\}\sin\theta, \tag{70}$$

$$p = -\frac{\eta}{R}(V_0 - 2d)\left(\frac{R}{r}\right)^2 \cos\theta. \tag{71}$$

Similarly, for the radial component of the velocity of the internal fluid (64) we have

$$V_r'(R,\theta) = 2\left\{\frac{a'}{10} + c'\right\}\cos\theta = 0, \tag{72}$$

Hence

$$a' = -10c'. \tag{73}$$

Then the system of equations for the internal fluid can be rewritten as

$$V_r' = 2c'\left\{1 - \left(\frac{r}{R}\right)^2\right\}\cos\theta, \tag{74}$$

$$V_\theta' = 2c'\left\{2\left(\frac{r}{R}\right)^2 - 1\right\}\sin\theta. \tag{75}$$

$$p' = -\frac{20c'\eta' r}{R^2}\cos\theta. \tag{76}$$

Conditions Eqs. (7)-(8) imposed in Eq. (69) and (74) on the surface of the drop give

$$\left.\frac{\partial V_r}{\partial \theta}\right|_{r=R} = \left.\frac{\partial V_r'}{\partial \theta}\right|_{r=R} = 0. \tag{77}$$

Therefore, the condition of stitching stresses $\sigma_{r\theta}$ giving by Eq. (13) is simplified

$$\eta\left(\left.\frac{\partial V_\theta}{\partial r}\right|_{r=R} - \frac{V_\theta(R)}{R}\right) = \eta'\left(\left.\frac{\partial V_\theta'}{\partial r}\right|_{r=R} - \frac{V_\theta'(R)}{R}\right). \tag{78}$$

Next, we have

$$V_\theta(R) = -\left\{\frac{1}{2}V_0 + 2d\right\}\sin\theta, \tag{79}$$

$$\frac{\partial V_\theta}{\partial r} = \left\{-V_0\left(\frac{R}{2r^2}\right) + d\left[\frac{R}{r^2} + \frac{3}{R}\left(\frac{R}{r}\right)^4\right]\right\}\sin\theta, \tag{80}$$

$$\left.\frac{\partial V_r}{\partial\theta}\right|_{r=R} = \left\{-V_0\frac{1}{2R} + d\frac{4}{R}\right\}\sin\theta = \frac{8d - V_0}{2R}\sin\theta, \tag{81}$$

$$\left.\frac{\partial V_r}{\partial\theta}\right|_{r=R} - \frac{V_\theta(R)}{R} = \frac{8d - V_0}{2R}\sin\theta + \frac{1}{R}\left\{\frac{1}{2}V_0 + 2d\right\}\sin\theta = \frac{6d}{R}\sin\theta. \tag{82}$$

Similarly, we find

$$V_\theta'(R) = 2c'\sin\theta, \tag{83}$$

$$\frac{\partial V_\theta'}{\partial r} = 2c'\frac{1}{R}\left\{4\frac{r}{R}\right\}\sin\theta, \tag{84}$$

$$\left.\frac{\partial V_\theta'}{\partial r}\right|_{r=R} = 8c'\frac{1}{R}\sin\theta, \tag{85}$$

$$\left.\frac{\partial V_\theta'}{\partial r}\right|_{r=R} - \frac{V_\theta'(R)}{R} = \frac{6c'}{R}\sin\theta, \tag{86}$$

Substituting Eq. (82) and Eq. (86) into Eq. (78), we obtain

$$c' = \frac{\eta}{\eta'}d. \tag{87}$$

Then

$$V_r' = 2\frac{\eta}{\eta'}d\left\{1 - \left(\frac{r}{R}\right)^2\right\}\cos\theta, \tag{88}$$

$$V_\theta' = 2\frac{\eta}{\eta'}d\left\{2\left(\frac{r}{R}\right)^2 - 1\right\}\sin\theta. \tag{89}$$

$$p' = -\frac{\eta}{\eta'}d\frac{20\eta'r}{R^2}\cos\theta = -d\frac{20\eta r}{R^2}\cos\theta. \tag{90}$$

Let's calculate the expression

$$-p(R) + 2\eta\left.\frac{\partial V_r}{\partial r}\right|_R = \frac{3\eta}{R}(V_0 + 2d)\cos\theta. \tag{91}$$

Similarly, we find

$$-p'(R) + 2\eta' \frac{\partial V_r'}{\partial r}\bigg|_{r=R} = d\frac{20\eta r}{R^2}\cos\theta - \frac{8\eta'\eta}{\eta' R^2}d\cos\theta = \frac{12 d\eta}{R}\cos\theta. \tag{92}$$

Substituting Eq. (91) and Eq. (92) into the boundary condition Eq. (14), we obtain

$$V_0 = 2d + \frac{(\rho-\rho')gR^2}{3\eta} \tag{93}$$

The remaining boundary condition Eq. (56), taking into account conditions (7)-(8), has the form

$$-p(R) + 2\eta\frac{V_\theta(R)\cot\theta}{R} + p'(R) - 2\eta'\frac{V_\theta'(R)\cot\theta}{R} = (\rho-\rho')gR\cos\theta. \tag{94}$$

Hence

$$d = -\frac{(\rho-\rho')gR^2}{30\eta}. \tag{95}$$

Substituting Eq. (95) into Eq. (93), one can obtain

$$V_0 = \frac{4(\rho-\rho')gR^2}{15\eta}. \tag{96}$$

Thus, replacing the condition for stitching the velocities of two liquids, internal and external, allows us to take into account the remaining unconsidered conditions for stitching stresses. The velocity (96) exceeds the velocity of the solid sphere, determined by the Stokes drag force formula (49), by ⅘ times. At the same time, there is no dependence on the viscosity of the inner sphere, which qualitatively distinguishes formula (96) from the Hadamard-Rybczynski Eq. (47).

The remaining undefined coefficients (68), (87) and (73) take the form

$$b = \frac{(\rho-\rho')gR^2}{3\eta}, \quad c' = -\frac{(\rho-\rho')gR^2}{30\eta'}, \quad a' = \frac{(\rho-\rho')gR^2}{3\eta'}. \tag{97}$$

Let us write out the solutions for the external and internal liquids by substituting Eq. (95) in Eqs. (69)-(71) and Eqs. (88)-(90), respectively.

1) The external liquid

$$V_r = \left\{4 - \frac{5R}{r} + \left(\frac{R}{r}\right)^3\right\}\frac{(\rho-\rho')gR^2}{15\eta}\cos\theta, \tag{98}$$

$$V_\theta = \left\{5\frac{R}{r} + \left(\frac{R}{r}\right)^3 - 8\right\}\frac{(\rho-\rho')gR^2}{30\eta}\sin\theta, \tag{99}$$

$$p = -\frac{(\rho-\rho')gR}{3}\left(\frac{R}{r}\right)^2\cos\theta. \tag{100}$$

Stream function is (Table 1):

$$\psi = R^2 \sin\theta \left\{ \frac{b}{2}\left(\frac{R}{r}\right)^{-1} + d\left(\frac{R}{r}\right) \right\} P_1^1 \tag{101}$$

or

$$\psi = \frac{(\rho-\rho')gR^4}{6\eta}\sin^2\theta \left\{ \left(\frac{R}{r}\right)^{-1} - \frac{R}{5r} \right\}. \tag{102}$$

2) The internal liquid

$$V_r' = -\frac{\eta}{\eta'}\frac{(\rho-\rho')gR^2}{15\eta}\left\{1-\left(\frac{r}{R}\right)^2\right\}\cos\theta, \tag{103}$$

$$V_\theta' = -\frac{\eta}{\eta'}\frac{(\rho-\rho')gR^2}{15\eta}\left\{2\left(\frac{r}{R}\right)^2 - 1\right\}\sin\theta. \tag{104}$$

$$p' = \frac{2(\rho-\rho')gr}{3}\cos\theta. \tag{105}$$

Stream function is (Table 1):

$$\psi' = -R^2 \sin\theta \left\{ \frac{a'}{10}\left(\frac{r}{R}\right)^4 + c'\left(\frac{r}{R}\right)^2 \right\} P_1^1. \tag{106}$$

or

$$\psi' = \frac{(\rho-\rho')gR^4}{30\eta'}\sin\theta \left\{ \left(\frac{r}{R}\right)^2 - \left(\frac{r}{R}\right)^4 \right\} P_1^1. \tag{107}$$

On the surface of a drop, the tangential components of the velocities are determined by the formulas

$$V_\theta(R,\theta) = -\frac{(\rho-\rho')gR^2}{15\eta}\sin\theta, \qquad V_\theta'(R,\theta) = -\frac{(\rho-\rho')gR^2}{15\eta'}\sin\theta, \tag{108}$$

Then relative slip velocity is:

$$\Delta V_\theta(R,\theta) = V_\theta(R,\theta) - V_\theta'(R,\theta) = \frac{(\rho-\rho')gR^2}{15}\left(\frac{1}{\eta'} - \frac{1}{\eta}\right)\sin\theta. \tag{109}$$

Eq. (109) shows that if the viscosities of the two liquids are the same, the slip velocity (109) vanishes, so we return to the boundary condition of the Hadamard and Rybczynski model (9).

Similar to the Navier boundary condition used to describe the slip of a liquid on a solid surface, we can introduce an effective slip length for a liquid drop in an external liquid:

$$\sigma_{\theta r}(\theta) = \eta\left(\frac{1}{r}\frac{\partial V_r}{\partial \theta} + \frac{\partial V_\theta}{\partial r} - \frac{V_\theta}{r}\right)_{r=R} = \frac{\eta}{\lambda}(V_\theta(R,\theta) - V_\theta'(R,\theta)). \tag{110}$$

According to equation (78), we have

$$\sigma_{\theta r}(\theta) = \sigma_{\theta r}{}'(\theta) = \eta\left(\left.\frac{\partial V_\theta}{\partial r}\right|_{r=R} - \frac{V_\theta(R)}{R}\right) = \eta'\left(\left.\frac{\partial V_\theta{}'}{\partial r}\right|_{r=R} - \frac{V_\theta{}'(R)}{R}\right), \quad (111)$$

where

$$\left.\frac{\partial V_\theta}{\partial r}\right|_{r=R} = -4\frac{(\rho-\rho')gR}{15\eta}\sin\theta. \quad (112)$$

Let us calculate the slip length $\lambda$ from the equation

$$\sigma_{\theta r}(\theta) = \eta\left(\left.\frac{\partial V_\theta}{\partial r}\right|_{r=R} - \frac{V_\theta(R,\theta)}{R}\right) = \frac{\eta}{\lambda}(V_\theta(R,\theta) - V_\theta{}'(R,\theta)), \quad (113)$$

$$-\frac{(\rho-\rho')gR}{5}\sin\theta = \frac{\eta}{\lambda}\frac{(\rho-\rho')gR^2}{15}\left(\frac{1}{\eta'} - \frac{1}{\eta}\right)\sin\theta. \quad (114)$$

or

$$\lambda = \frac{R}{3}\left(1 - \frac{\eta}{\eta'}\right). \quad (115)$$

Similarly, for the internal fluid, the slip length $\lambda'$ is given by the relation

$$\sigma_{\theta r}{}'(\theta) = \eta'\left(\left.\frac{\partial V_\theta{}'}{\partial r}\right|_{r=R} - \frac{V_\theta{}'(R,\theta)}{R}\right) = \frac{\eta'}{\lambda'}(V_\theta{}'(R,\theta) - V_\theta(R,\theta)), \quad (116)$$

hence

$$\lambda' = \frac{R}{3}\left(1 - \frac{\eta'}{\eta}\right). \quad (117)$$

We see that if the viscosities of both liquids are equal, $\eta = \eta'$, then the slip length becomes zero, and a boundary condition is no-slip. In this case, Eqs. (47) and (96) become identical, i.e. the Hadamard and Rybczynski model is matched with the continuous viscous stress tensor model described in this section.

### 4. Model with arbitrary slip length at liquid-liquid interface

Experiments show, in most cases for quite small droplets, that the Stokes drag force formula obtained for a rigid sphere works, and for larger droplets, the velocity approaches the description by the Hadamard-Rybczynski equation, although there are intermediate variants. Therefore, neither the Hadamard-Rybczynski model nor the continuous viscous stress tensor model are universal for describing the phenomena.

Obviously, a more universal description can be achieved by replacing the boundary condition (9) in the Hadamard-Rybczynski model with the boundary condition of partial slip (110) with an arbitrary slip length λ:

$$\sigma_{\theta r}(\theta) = \eta \left( \frac{1}{r}\frac{\partial V_r}{\partial \theta} + \frac{\partial V_\theta}{\partial r} - \frac{V_\theta}{r} \right)_{r=R} = \frac{\eta}{\lambda}(V_\theta(R,\theta) - V_\theta'(R,\theta)). \tag{118}$$

In this case, all other boundary conditions of the Hadamard-Rybczynski model should remain in force, and the conditions of the continuous viscous stress tensor model (53)-(54) will have to be abandoned, since they are generally not confirmed by experiment. In this case, the number of boundary conditions is fixed by the mathematical model.

Thus, let us return to the conclusion drawn in the previous section and replace condition (56), which is introduced after formula (93), with condition (118). As shown in the previous section, condition (118) can be rewritten as:

$$\lambda \left( \frac{\partial V_\theta}{\partial r}\bigg|_{r=R} - \frac{V_\theta(R,\theta)}{R} \right) = V_\theta(R,\theta) - V_\theta'(R,\theta). \tag{119}$$

At λ=0, condition (119) becomes condition (9), which corresponds to the Hadamard-Rybczynski model. Substituting into (119) the projections of the velocities $V_\theta$ and $V_\theta'$, determined by Eqs. (70) and (89), respectively, as well as Eq. (82), we find an equation that allows us to calculate the coefficient $d$:

$$\lambda \frac{6d}{R} \sin\theta = \left( -\left\{\frac{1}{2}V_0 + 2d\right\} - 2\frac{\eta}{\eta'}d \right)\sin\theta, \tag{120}$$

or

$$d = -\frac{(\rho-\rho')gR^2}{6\eta}\left( \frac{6\lambda}{R} + 3 + \frac{2\eta}{\eta'} \right)^{-1}. \tag{121}$$

Substituting Eq. (121) into Eq. (93), we obtain

$$V_0 = \frac{(\rho-\rho')gR^2}{3\eta}\left\{ 1 - \left( \frac{6\lambda}{R} + 3 + \frac{2\eta}{\eta'} \right)^{-1} \right\} \tag{122}$$

or

$$V_0 = \frac{2(\rho-\rho')gR^2}{3\eta}\frac{1+\eta\eta'^{-1}+3\lambda R^{-1}}{3+2\eta\eta'^{-1}+6\lambda R^{-1}} \tag{123}$$

or

$$V_0 = \frac{2(\rho-\rho')gR^2}{3\eta}\frac{\eta+\eta'(1+3\lambda R^{-1})}{2\eta+3\eta'(1+2\lambda R^{-1})}. \tag{124}$$

At λ=0, we have the no-slip condition, under which Eq. (124) becomes the Hadamard-Rybczynski Eq. (47).

In the limit of infinitely high viscosity of the internal fluid for a limited value of λ, we have

$$V_0 = \frac{2(\rho-\rho')gR^2}{9\eta}\frac{1+3\lambda R^{-1}}{1+2\lambda R^{-1}}. \tag{125}$$

Formula (125) describes the sliding of a solid ball in an external fluid with viscosity η and slip length λ. It coincides with the previously published formula [8,9].

Thus, the resulting formula generalizes the Hadamard-Rybczynski formula and its limit – the Stokes formula by introducing partial slip.

The remaining undefined coefficients (68), (87) and (73) take the form

$$b = \frac{(\rho-\rho')gR^2}{3\eta}. \tag{126}$$

$$c' = -\frac{(\rho-\rho')gR^2}{6\eta'}\left(\frac{6\lambda}{R}+3+\frac{2\eta}{\eta'}\right)^{-1}. \tag{127}$$

$$a' = \frac{5(\rho-\rho')gR^2}{3\eta'}\left(\frac{6\lambda}{R}+3+\frac{2\eta}{\eta'}\right)^{-1}. \tag{128}$$

Let us write out the solutions for the external and internal liquids by substituting Eq. (95) in Eqs. (69)-(71) and Eqs. (88)-(90), respectively.

1) The external liquid

$$V_r = \frac{(\rho-\rho')gR^2}{3\eta}\left\{1-\frac{R}{r}-\left(\frac{6\lambda}{R}+3+\frac{2\eta}{\eta'}\right)^{-1}\left[1-\left(\frac{R}{r}\right)^3\right]\right\}\cos\theta, \tag{129}$$

$$V_\theta = \frac{(\rho-\rho')gR^2}{6\eta}\left\{\frac{R}{r}-2+\left(\frac{6\lambda}{R}+3+\frac{2\eta}{\eta'}\right)^{-1}\left[2+\left(\frac{R}{r}\right)^3\right]\right\}\sin\theta, \tag{130}$$

$$p = -\frac{(\rho-\rho')gR^3}{3r^2}\cos\theta. \tag{131}$$

Stream function is (Table 1):

$$\psi = \sin\theta\frac{(\rho-\rho')gR^4}{6\eta}\left\{\left(\frac{R}{r}\right)^{-1}-\left(\frac{6\lambda}{R}+3+\frac{2\eta}{\eta'}\right)^{-1}\left(\frac{R}{r}\right)\right\}P_1^1 \tag{132}$$

2) The internal liquid

$$V_r' = -\frac{(\rho-\rho')gR^2}{3\eta'}\left(\frac{6\lambda}{R}+3+\frac{2\eta}{\eta'}\right)^{-1}\left\{1-\left(\frac{r}{R}\right)^2\right\}\cos\theta, \tag{133}$$

$$V_\theta' = -\frac{(\rho-\rho')gR^2}{3\eta'}\left(\frac{6\lambda}{R}+3+\frac{2\eta}{\eta'}\right)^{-1}\left\{2\left(\frac{r}{R}\right)^2-1\right\}\sin\theta, \tag{134}$$

$$p' = \frac{10(\rho-\rho')gr}{3}\left(\frac{6\lambda}{R}+3+\frac{2\eta}{\eta'}\right)^{-1}\cos\theta. \tag{135}$$

Stream function is (Table 1):

$$\psi' = -\sin\theta\frac{(\rho-\rho')gR^4}{6\eta'}\left(\frac{6\lambda}{R}+3+\frac{2\eta}{\eta'}\right)^{-1}\left\{\left(\frac{r}{R}\right)^4-\left(\frac{r}{R}\right)^2\right\}P_1^1. \tag{136}$$

## 5. Discussion

The Hadamard-Rybczynski equation is usually used to describe the motion of a small drop of liquid in another liquid. In its derivation, the no-slip boundary condition at the liquid-liquid interface was used. This somewhat contradicts the initial assumption that both liquids are immiscible (poorly soluble in each other). Therefore, a more natural and generalizing condition is the partial slip of one liquid on the surface of the other (118), which is a modification of the Navier condition originally introduced for the liquid-solid interface.

In this paper, the Navier condition is applied to the liquid-liquid boundary for the first time. A generalized Hadamard-Rybczynski equation (124) is obtained, which at $\lambda=0$ transforms into the usual Hadamard-Rybczynski equation (47). At $\lambda$ defined by formula (115), we arrive at a model with a slip length controlled by the continuity of the components of the viscous stress tensor at the interface of two fluids. For infinite viscosity of the drop, Eq. (124) becomes a well-known relation generalizing the Stokes drag force for a solid sphere, taking into account the boundary condition of partial slip Eq. (125), obtained using the Navier boundary condition [8,9].

The equation (124), which can be rewritten as

$$V_0 = \frac{2(\rho-\rho')gR^2}{3\eta}\frac{\eta+\eta'+3\lambda\eta' R^{-1}}{2\eta+3\eta'+6\lambda\eta' R^{-1}}. \tag{137}$$

of course, does not exhaust all possible variants of boundary conditions, and therefore cannot claim universality, but there is a fairly wide and important class of substances for the description of which this equation can be applied. From a mathematical point of view, it is obvious that this is approximately the range of substances for the description of which the Boussinesq equation can be used [1,10]:

$$V_0 = \frac{2(\rho-\rho')gR^2}{3\eta}\frac{\eta+\eta'+2e(3R)^{-1}}{2\eta+3\eta'+2eR^{-1}}, \tag{138}$$

where $e$ is the "coefficient of surface viscosity" introduced by Boussinesq. Indeed, the structure of expression (138) almost exactly coincides with the expression (137) we obtained. In this case, the concept of the slip length seems to make more physical sense.

There are numerous experiments [1,3] that demonstrate the influence of surfactants on the rate of fall or rise of liquid droplets. These experiments can obviously be naturally understood taking into account that surfactants undoubtedly affect the slip length.

In general, it should be noted that the hydrodynamic approach (Navier-Stokes equations) is based primarily on viscosity, which is directly included in these equations. In this sense, both the Boussinesq approach and the approach with a non-zero slip length are more natural precisely within the framework of the hydrodynamic approach.

At the same time, of course, one cannot deny the existence in nature of other, more specific mechanisms that are associated with the uneven environment of the liquid-liquid interface by surfactants [1,3].

The field of application of Eq.(124) is the motion of hydrophobic (lipophilic) liquids in water and vice versa, i.e. the description of stratification and sedimentation of aqueous emulsions or vice versa – water droplets in oils, etc. Presumably, the best applicability of this equation should be expected for the liquid-liquid interface of immiscible liquids with not very different viscosity and density (water – hydrocarbons, water – higher alcohols, in general: aqueous emulsions, water – lipophilic organic liquids and oils, etc.). The slip length allows the experiment and theory to be reconciled.

**References**

1. Levich V.G. Physicochemical hydrodynamics. Prentice-Hall, Inc. Englewood Cliffs, N.J. 1962. 700 p.

2. Silvey O.W. The Fall of Mercury Droplets in a Viscous Medium. Phys. Rev., 7, 106 (1916)

3. Ervik A., Bjorklund E. The transition in settling velocity of surfactant-covered droplets from the Stokes to the Hadamard–Rybczynski solution. European Journal of Mechanics / B Fluids, 2017, 66, pp. 10-19.

4. Landau L.D., Lifshitz E.M., Fluid Mechanics: Landau and Lifshitz: Course of Theoretical Physics. Vol. 6 (Pergamon Press, Oxford, 1987).

5. Batchelor G.K., An Introduction to Fluid Dynamics (Cambridge University Press, Cambridge, 2000).

6. Lebedev-Stepanov P. Appendix in: ArXiv:2411.15853 [physics.flu-dyn].

7. Arfken G.B., Weber H.-J., Harris F. E., Mathematical methods for physicists: a comprehensive guide (Elsevier, 2012).

8. Boehnke U.C., Remmler T., Motschmann H., Wurlitzer S., Hauwede J., Fischer M. Th. Partial air wetting on solvophobic surfaces in polar liquids, J. Colloid Int. Sci. 211, 243 (1999).

9. Lauga E., Brenner M.P., Stone H.A. Microfluidics: the no-slip boundary condition. In: J. Foss, C. Tropea, A.L. Yarin (Eds.) Handbook of Experimental Fluid Dynamics (Springer, New York, 2007).

10. Boussinesq, J., Comptes Rendus des Seances de l'Academie des Sciences, 156, 1124 (1913) [http://gallica.bnf.fr/ark:/12148/bpt6k3109m.image.f1124.langFR].